\documentclass{article}
\usepackage{quel-art}
\usepackage[12pt]{xy}
\usepackage{graphicx}
\usepackage{amssymb}
\usepackage{colordvi}
\volnumber{20} \issuenumber{1} \edyear{2004}                             
\frompage{000} \topage{000}                                              
\recrevdate{1 August 2006}                                              
\def\rightmark{Witnessing Entanglement}
\def\leftmark{M. Stobi\'nska and K. W\'odkiewicz}

\title{Witnessing Entanglement with Second-Order Interference\\ and
  Stokes Parameters}
\authors{
{Magdalena Stobi\'nska$^1$ and Krzysztof W\'odkiewicz$^{1,2}$ %
\index{Stobi\'nska, M.} 
\index{W\'odkiewicz, K.} 
}\\[2.812mm]
{\normalsize
\hspace*{-8pt}$^1$ Instytut Fizyki Teoretycznej, Uniwersytet Warszawski,\\
00-681 Warszawa, Poland\\[0.2ex]
\hspace*{-8pt}$^2$ Department of Physics and Astronomy, University of
New Mexico,\\
Albuquerque, NM 87131-1156, USA\\[0.2ex]
}}

\abstract{We propose to use Stokes parameter as an entanglement
  witness for correlated EPR mixed states of light.  Such states can
  be generated with a beam splitter acting on two mixed squeezed
  states of light. Stokes witness operators are closely related to the
  Hanbury-Brown and Twiss interference and can be used to test
  entanglement in balanced homodyne experiments involving fluctuations
  of quantum quadratures of the electric field.} 


\makeindex
\begin{document}

\maketitle

\section{Introduction}

In the famous paper questioning the reality of quantum mechanics,
Einstein, Podolsky and Rosen \cite{EPR} introduced a quantum state
with nonclassical correlations between two spatially separated massive
particles. In its original formulation this EPR correlated state was
presented in the following general form
\begin{equation}
\Psi(x_1,x_2)=\sum_{n=1}^{\infty}\psi_n(x_1)u_n(x_2).
\label{schmidt}
\end{equation}
In this remarkable formula we recognize the Schmidt decomposition of
an entangled state. The EPR understanding of the probabilistic
interpretation of quantum mechanics has been criticized by N. Bohr
\cite{Bohr}. This famous debate about the foundations of quantum
theory and the meaning of entanglement has drawn the attention of many
physicists.

As an example of a correlated state given by Eq.(\ref{schmidt}), EPR
have used the following wave function
\begin{equation}
\Psi(x_1,x_2) \simeq \delta(x_1-x_2).
\label{eq:delta_x}
\end{equation}
This wave function reflects the presence of perfect correlations
between the coordinates of the two particles. If the first particle is
at $x_1$, the other one is at $x_1$ with certainty. Similarly, there
exists a perfect correlation in the momenta of the two particles
\begin{equation}
\widetilde{\Psi}(p_2,p_2) \simeq \delta(p_1 + p_2).
\label{eq:delta_p}
\end{equation}

The most important tools used in tests of quantum entanglement have
been based on various Bell inequalities. These inequalities became a
driving force behind a very active experimental branch of a quantum
optics \cite{Aspect1982,Zeilinger1998,Geneva1998,Grangier2004}. Most
of the experimental tests have been limited to entangled states of two
qubits.  For entangled systems described by continuous variables like
the EPR wave function, Bell inequalities of some kind are violated as
it has been demonstrated recently \cite{BaWo98}.

The entangled states defined by the Schmidt decomposition are systems
described by wave functions, i.e., are pure quantum states.  For a
quantum system described by a mixed statistical operator $\rho$, it
turns out that the concept of quantum entanglement, as defined by
(\ref{schmidt}), has to be generalized. In the general case of a
density operator, rather than a wave function, one uses the definition
of quantum separability introduced by Werner \cite{Werner89}. A
general quantum density operator of a two-party system is separable if
it is a convex sum of product states,
\begin{equation}
 \rho = \sum_k p_k\,\rho_1^{(k)}\otimes\rho_2^{(k)}\quad
\mathrm{with} \quad \sum_k p_k=1 \quad \mathrm{and} \quad p_k>0\,,
\label{separability}
\end{equation}
where $\rho_1^{(k)}$ and $\rho_2^{(k)}$ are statistical operators of
the two subsystems in question. If this criterion is not satisfied the
state is called non-separable or simply entangled.

The problem of establishing which mixed state is separable and which
is not, is much more complex, and the question of experimental tests
for such systems is still an open problem.  Therefore, within the last
decade a notion of an entanglement witness has generated a lot of
interest as a experimentally feasible way of entanglement test for
mixed states. The idea relies on constructing an operator which mean
value is positive for separable and negative for entangled states
\begin{displaymath}
 \langle \mathcal{W} \rangle = \left\{ \begin{array}{ll}
       \ge 0, & \rho\;  \mathrm{is} \; \mathrm{separable}, \\
       < 0, & \rho\;  \mathrm{is} \; \mathrm{entangled}.
\end{array} \right.
\end{displaymath}
Such an operator has been discussed first for discrete variables
\cite{Lewenstein2000}. A general method for the experimental detection
of entanglement for qubits is discussed in \cite{Guhne2002}. The other
experimentally feasible method relies on violating the entropy
inequalities \cite{Horodecki2003}.  Another entanglement witness for
continuous variables, based on the measurement of the fluctuations of
photocurrent difference between two correlated modes of detected
light, has been introduced in \cite{paris1999}.

A method relying on the Hanbury-Brown and Twiss interference, in which
one can introduce an entanglement witness operator
$\mathcal{W}^{(HBT)} $ has been presented in \cite{Stobinska2005}.

In this paper we explicitly construct an operator which witnesses
continuous variable entanglement of the EPR state. This operator can
be expressed in terms of quantum uncertainties of the Stokes
parameter. We show that the measurement of witness operator is
equivalent to homodyne detection of quantum correlations and
fluctuations of the electric field quadratures.  If the value of the
quantum uncertainty is below the shot noise the state is entangled.
Our work focuses on general mixed gaussian states, involving as an
example mixed EPR states.

\section{Thermal mixed squeezed states}

Let us start our discussion with a general one-mode mixed gaussian
state. A single mode electric field operator, oscillating with
frequency $\omega$, can be written in the following form
\begin{equation}
E = E_0 ( a e^{-i\omega t} + a^{\dagger} e^{i\omega t})
= \sqrt{2}E_0(X_1\cos\omega t + X_2\sin\omega t),
\label{efield}
\end{equation}
where $X_1 = \frac{a+a^{\dagger}}{\sqrt{2}}$ and $X_2=
\frac{a-a^{\dagger}}{\sqrt{2}i}$ are amplitude and phase hermitian
quadratures of the electric field.

A one-mode gaussian state is completely characterized by the elements
of its covariance matrix given by the following second-order moments
\begin{equation}
\langle a^{\dagger}a \rangle = \bar{n},\; \langle a^2 \rangle = -m,
\end{equation}
where $\bar{n}$ is a mean number of photons in the mode, and $m$ is a
squeezing parameter, which we assume for simplicity to be real.  We
have assumed that $\langle a \rangle=\langle a^{\dagger} \rangle=0$,
which can always be arranged with a suitable unitary shift of $a$ and
$a^{\dagger}$.

Knowing those two parameters it is easy to construct a positive
density operator for the state
\begin{equation}
\rho_a = \frac{ e^{-N\, a^{\dagger}a - M^* a^2 - M\,{a^{\dagger}}^2}}{Z},
\end{equation}
where $N$ and $M$ are known functions of $\bar{n}$ and $m$
\cite{GardinerZoller2000}. In this formula $Z$ is the partition
function. This mixed squeezed state leads to the following uncertainty
of electric field quadratures
\begin{equation}
\Delta X_1 = \sqrt{\bar{n}+\frac{1}{2}-m}, \;
\Delta X_2 = \sqrt{\bar{n}+\frac{1}{2}+m},
\end{equation}
and the Heisenberg uncertainty relation is
\begin{equation}
\Delta X_1 \Delta X_2 =
\sqrt{\left(\bar{n}+\frac{1}{2}\right)^2-m^2},
\end{equation}
which imposes the following condition for the squeezing parameter
\begin{equation}
0 \le  |m| \le \sqrt{\bar{n}(\bar{n}+1)}.
\end{equation}

It is clear that this condition guarantees that $\rho_a\geq 0$. The
notion that $m$ is the squeezing parameter of quantum quadratures is
obvious from these relations. Squeezing means in this case
fluctuations below the vacuum noise, i.e.,
\begin{equation}
\bar{n} < |m|.
\end{equation}

For $|m|= \sqrt{\bar{n}(\bar{n}+1)}$, this state reduces to the single
mode pure state.

\section{Entangling photons}

One of the simplest way to entangle photons is to use a beam splitter.
The action of the beam splitter is a nonlocal operation that may, in
general, produce correlated photons. The use of a beam splitter to
entangle squeezed light has been implemented in several experiments
\cite{Glockl2003,Huntington2005}.  This process is
described as a $50/50$ beam splitter (BS) transformation of a product
state into a correlated state
\begin{equation}\label{bs}
BS:\ \rho_a\otimes\rho_b \mapsto \rho_{cd}.
\end{equation}
The operators $\rho_{a,b}$ are the density operators of the single
mode mixed squeezed states with equal number of photons but opposite
squeezing phases
\begin{eqnarray}
\langle a^{\dagger}a\rangle &=& \langle b^{\dagger}b \rangle=\bar{n}, 
\nonumber\\
\langle a^2 \rangle &=& -\langle b^2 \rangle=-m.
\label{correlation}
\end{eqnarray}

The relation between input and output modes operators on a $50/50$
beam splitter have the well known form
\begin{equation}
c = \frac{a+b}{\sqrt{2}}, \; d = \frac{a-b}{\sqrt{2}}.
\label{outcoming_modes}
\end{equation}
The two output beams form a correlated squeezed mixed EPR state with
equal mean number of photons in each mode
\begin{equation}
\langle c^{\dagger}c\rangle = \langle d^{\dagger}d\rangle = \bar{n}
\label{EPR_n}
\end{equation}
and with the following correlation parameter
\begin{equation}
\langle cd\rangle =  \frac{\langle a^2\rangle 
- \langle b^2\rangle}{2}=-m.
\label{EPR_m}
\end{equation}
It is clear that the beam splitter correlates the two beams
\begin{equation}
BS:\ (\bar{n},m)_a\otimes(\bar{n},-m)_b \mapsto (\bar{n},m)_{cd}.
\end{equation}

For $|m|= \sqrt{\bar{n}(\bar{n}+1)}$ this state reduces to the well
known pure two-mode squeezed state, generated in a process of
nondegenerate optical parametric amplification (NOPA)
\begin{equation}
|\mathrm{NOPA}\rangle = \sum_{n=0}^{\infty} \sqrt{p_n}
|n,n\rangle,
\end{equation}
where $p_n = \frac{\bar{n}^n}{(1+\bar{n})^{n+1}}$ is a Bose-Einstein
photon distribution.  This state is the closest experimental
realization, of the original EPR state. In the limit of infinite
intensity, $\bar{n} \rightarrow \infty$, this state with $p_n\sim 1$
becomes the EPR state (\ref{eq:delta_x})
\begin{equation}
\lim_{\bar{n} \rightarrow \infty}\langle x_1,x_2|\mathrm{NOPA}\rangle 
= \sum_{n=0}^{\infty} \langle
x_1,x_2|n,n\rangle \sim \delta(x_1-x_2).
\end{equation}

\begin{figure}[h!]
\begin{center}
\scalebox{0.9}{\includegraphics{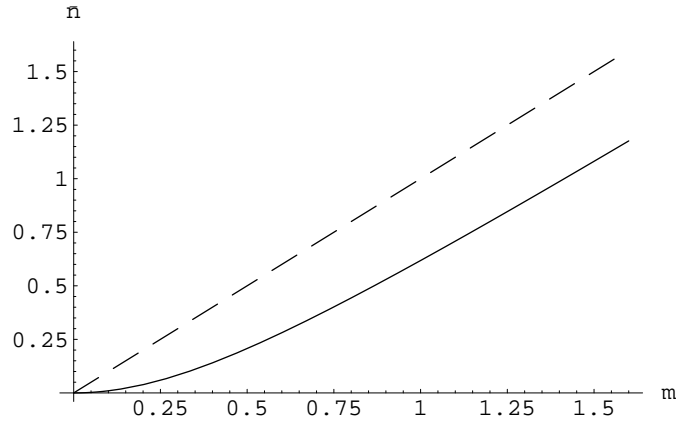}}
\caption{The values on and above the curve $\bar{n}
  = \frac{1}{2}\left(-1 + \sqrt{1+4m^2}\right)$ correspond to the
  physical states (with a positive density operator) - EPR mixed
  states.  The states between the curve and the dashed line
  $\bar{n}=m$ are entangled. The EPR states above the dashed line are
  classically correlated, so separable.}
\label{EPR_parameters}
\end{center}
\end{figure}

The separability condition for the two mode mixed state has been
addressed in several publications. In this paper we follow the results
obtained in \cite{Englert2003,Englert2002}. Figure
\ref{EPR_parameters} shows the results computed for the general mixed
EPR state.

\section{Witnessing entanglement}

In reference \cite{Stobinska2005} it has been shown that the
entanglement witness operator, related to the second-order
Hanbury-Brown and Twiss interference, is given by the following
hermitian operator
\begin{equation}
\label{eq:witness}
 \mathcal{W}^{(HBT)}=\frac{1}{2}- \frac{2 a^\dagger  b^\dagger a b
 + {b^{\dagger}}^2 a^2 + {a^{\dagger}}^2
  b^2 }{\langle :(I_a+I_b)^2:\rangle}.
\end{equation}
The mean value of this operator allows to witness the entanglement in
the outcoming modes $c$ and $d$. Using Eq. (\ref{outcoming_modes}) we
obtain
\begin{displaymath}
\mathrm{Tr}\{\mathcal{W}^{(HBT)}\rho\}=  \frac{\bar{n}^2
-|m|^2}{2(3\bar{n}^2 + |m|^2)} = \left\{ \begin{array}{ll}
       \ge 0, & \mathrm{separable}, \\
       < 0, & \mathrm{entangled}.
\end{array} \right.
\end{displaymath}

\section{Hong-Ou-Mandel interference}

The entanglement witness which has been introduced in the last section
is experimentally feasible quantity. It is directly related to the
second-order visibility $v^{(2)}$ observed in the Hong-Ou-Mandel time
resolved interference.

The joint coincidence of detecting photons from mode $c$ and $d$ at
delayed times $t$ and $t+\tau$ can be obtained directly from the
following second-order temporal coherence function
\begin{equation}
G^{(2)}(t, t+\tau) = \left\langle E^{(-)}_c(t)\, E^{(-)}_d(t+\tau)\,
E^{(+)}_d(t+\tau)\, E^{(+)}_c(t)\right\rangle.
\end{equation}

Assuming the random phases in the output state and random phases
typical for a stationary stochastic phase diffusion model in mode
functions, we achieve the probability of the joint detection depending
only on the delay~\cite{Stobinska2006}
\begin{equation}
p(\tau)= 1 - v^{(2)}\, \exp\left\{-\left(\frac{\tau}{\tau_c}\right)^2\right\},
\end{equation}
where $\tau_c$ is a coherence time. For the EPR correlated state, the
visibility of interference fringes is equal to
\begin{equation}
v^{(2)} = \frac{\bar{n}^2 + |m|^2}{3\bar{n}^2 + |m|^2}.
\label{eq:visibility_epr}
\end{equation}
The visibility is directly related to entanglement witness
\begin{equation}
\label{eq:witness_v}
 \langle \mathcal{W}^{(HBT)} \rangle = \frac{1}{2} - v^{(2)}.
\label{eq:witness_v}
\end{equation}
If the EPR state is separable, $\bar{n} \ge |m|$, the visibility is
$v^{(2)} \le \frac{1}{2}$ and the witness (\ref{eq:witness_v}) is
positive. For entangled states $v^{(2)} > \frac{1}{2}$, the witness
takes a negative value.

\section{Stokes Parameters}

The Stokes operators for the outgoing modes $c$ and $d$ are given by
the following expressions
\begin{eqnarray}
S_x &=& \frac{c^{\dagger}d + d^{\dagger}c}{2}, \;
S_z = \frac{c^{\dagger}c - d^{\dagger}d}{2},\\
S_y &=& \frac{c^{\dagger}d - d^{\dagger}c}{2i}, \;
S_0 = \frac{c^{\dagger}c + d^{\dagger}d}{2}.
\label{stokes}
\end{eqnarray}
These definitions of the Stokes operators are known as the Schwinger
representation of the angular momentum operators in terms of
annihilation operators of two harmonic oscillators.  These operators
provide a very useful and efficient tool in the theoretical and
experimental description of polarization entangled states
\cite{Heersink2003}. In this paper we will use the Stokes parameters
for continuous variable entangled squeezed states.

The Stokes operators form a set of noncommuting operators
\begin{equation}
[S_x,S_y] = iS_z
\end{equation}
and their joint measurement is limited by the Heisenberg uncertainty
relation
\begin{equation}
(\Delta S_y)^2(\Delta S_z)^2 \ge |\langle S_x \rangle|^2,
\end{equation}
which is equivalent to a positive density matrix condition
$\bar{n}(\bar{n}+1) \ge |m|^2$.

The entanglement witness (\ref{eq:witness}) can also be expressed in
terms of the normally ordered Stokes parameters
\begin{eqnarray}
 \langle \mathcal{W}^{(HBT)} \rangle &=& \frac{1}{2} - \frac{ \langle :S_x^2:
 \rangle}{\langle :S_x^2: \rangle + \langle :S_y^2: \rangle + \langle
 :S_z^2: \rangle} \\
&=& \frac{1}{2} - \frac{ \langle S_x^2
 \rangle - \frac{1}{2} \langle S_0
 \rangle }{\langle S_x^2 \rangle + \langle S_y^2 \rangle + \langle
 S_z^2 \rangle - \frac{3}{2} \langle S_0
 \rangle}.
\end{eqnarray}

The mean value of the witness takes a negative value if any of the
following inequalities is obeyed
\begin{eqnarray}
\langle :S_x^2: \rangle &>& \langle :S_y^2: \rangle + \langle :S_z^2:
\rangle,\\
\langle S_x^2 \rangle &>& \langle S_y^2 \rangle + \langle S_z^2
\rangle - \frac{1}{2} \langle S_0 \rangle.
\end{eqnarray}

\begin{figure}[h]
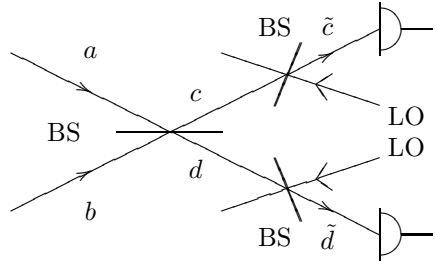

\begin{center}
  \begin{displaymath}
    \xy<0.7cm,0cm>:
    (2,0);(4,0)**@{-},
         (0,1.5);(7,-1.95)**@{-},
     (1.5,0.75)*\dir{>},
     (6.1,-1.5)*\dir{>},
         (0,-1.5);(7,1.95)**@{-},
     (6.1,1.5)*\dir{>},
     (1.5,-0.75)*\dir{>},
    (1.5,1.5)*{a}, (1.5,-1.5)*{b},
     (5,0.5);(5.5,1.7)**@{-},
     (5.02,0.5);(5.52,1.7)**@{-},
     (5,-0.5);(5.5,-1.7)**@{-},
     (5.02,-0.5);(5.52,-1.7)**@{-},
      (7.5,-0.3)*{\mbox{LO}},
      (7.5,0.3)*{\mbox{LO}},
      (4,1.5);(7,0.5)**@{-},
      (4,-1.5);(7,-0.5)**@{-},
     (6,0.6);(5.8,0.9)**@{-},
     (6.1,1.1);(5.8,0.9)**@{-},
     (6,-0.6);(5.8,-0.9)**@{-},
     (6.1,-1.1);(5.8,-0.9)**@{-},
%
    (7,-2.45);(7,-1.45)**@{-}, (7,-1.95)*\cir(0.5,0.5){r^l},
    (7,2.45);(7,1.45)**@{-}, (7,1.95)*\cir(0.5,0.5){r^l},
    (7.4,-1.95);(8,-1.95)**@{-},
    (7.4,1.95);(8,1.95)**@{-},
    (6,-2)*{\tilde{d}},
    (6,2)*{\tilde{c}},
    (3.5,0.7)*{c},
    (3.5,-0.7)*{d},
    (1,0)*{\mbox{BS}},
    (5,2)*{\mbox{BS}},
    (5,-2)*{\mbox{BS}},
   \endxy
  \end{displaymath}
\end{center}
\caption{Balanced homodyne detection. The outgoing modes, $c$ and $d$
  given by (\ref{EPR_n}) and (\ref{EPR_m}), are combined with a strong
  local oscillator beam at a $50/50$ beam splitter.}
\label{Fig:bhd}
\end{figure}

As we shall see below, the measurement of the uncertainty of a single
Stokes parameter can also constitute an entanglement witness. The
measurement of $\Delta S_z$ seems to be the best choice for an
experimental realization. This measurement can be done using balanced
homodyne detection with a strong local oscillator beam $\alpha =
|\alpha|e^{i\varphi}$. In Fig. \ref{Fig:bhd}, we have depicted an
experimental scheme of such a measurement. The two incoming modes $a$
and $b$ (\ref{correlation}) are correlated at a $50/50$ beam splitter.
The detectors measure intensities of the new displaced modes
$\tilde{c}$ and $\tilde{d}$. The Stokes parameter $S_z$ for the
detected outgoing modes $\tilde{c}$ and $\tilde{d}$ is given by
\begin{equation}
S_z = \frac{\tilde{c}^{\dagger}\tilde{c} -
  \tilde{d}^{\dagger}\tilde{d}}{2},
\end{equation}
where the modes $\tilde{c}$ and $\tilde{d}$ are displaced by the
strong pump
\begin{equation}
\tilde{c} = c + \alpha, \; \tilde{d} = d + \alpha.
\end{equation}

\noindent
Defining rotated quadratures as
\begin{equation}
X_c(\varphi) = \frac{c \, e^{-i\varphi} +
  c^{\dagger}e^{i\varphi}}{\sqrt{2}}, \;
X_d(\varphi) = \frac{d \, e^{-i\varphi} +
  d^{\dagger}e^{i\varphi}}{\sqrt{2}},
\end{equation}
the uncertainty of Stokes parameter $S_z$ is equal to
\begin{equation}
(\Delta S_z)^2 = \frac{\bar{n}(\bar{n}+1)-m^2}{2} + \frac{|\alpha|^2}{2} \left
  \langle (X_c(\varphi) - X_d(\varphi))^2 \right \rangle.
\end{equation}
The above formula simplifies to
\begin{equation}
(\Delta S_z)^2 = \frac{\bar{n}(\bar{n}+1)-m^2}{2} + \frac{|\alpha|^2}{2} \left
  (1 + 2(\bar{n} + m \cos 2\varphi) \right),
\label{delta_S_z}
\end{equation}
where $\langle X_c(\varphi) X_d(\varphi) \rangle = m \cos 2\varphi$ is
the correlation between the two quadratures.  Choosing $\varphi =
\frac{\pi}{2}$ and keeping quadratic terms in $\alpha$ only, the
formula (\ref{delta_S_z}) reduces~to
\begin{equation}
(\Delta S_z)^2 \sim \frac{|\alpha|^2}{2}(1+2(\bar{n}-m)). 
\label{shot_noise}
\end{equation}
The above expression shows that if the EPR state is entangled, so if $m
> \bar{n}$ the uncertainty of $S_z$ is below the shot noise
\begin{equation}
(\Delta S_z)^2 < \frac{|\alpha|^2}{2}.
\end{equation}
In a similar way other Stokes parameters can also serve as an
entanglement witness.

Evaluating the variance of the Stokes parameter $S_x$
\begin{equation}
S_x = \frac{\tilde{c}^{\dagger}\tilde{d} +
  \tilde{d}^{\dagger}\tilde{c}}{2},
\end{equation}
with $\langle S_x \rangle = |\alpha|^2$ we get
\begin{equation}
(\Delta S_x)^2 = \frac{\bar{n}(\bar{n}+1)+m^2}{2} + \frac{|\alpha|^2}{2} \left
  \langle (X_c(\varphi) + X_d(\varphi))^2 \right \rangle.
\end{equation}
In the strong field limit this formula simplifies to
\begin{equation}
(\Delta S_x)^2 \sim \frac{|\alpha|^2}{2} \left (1 + 2(\bar{n} 
- m \cos 2\varphi) \right).
\label{delta_S_x}
\end{equation}
Choosing $\varphi = 0$ and keeping quadratic terms in $\alpha$
only, the formula (\ref{delta_S_x}) reduces to formula
(\ref{shot_noise})\Red{.} The outcoming state is entangled if
$(\Delta S_x)^2$ is smaller than for a shot noise. Similarly for
$S_y$
\begin{equation}
S_y = \frac{\tilde{c}^{\dagger}\tilde{d} -
  \tilde{d}^{\dagger}\tilde{c}}{2}.
\end{equation}
Its mean value $\langle S_y \rangle = 0$ and the variance is equal to
\begin{equation}
(\Delta S_y)^2 = \frac{n(n+1)+m^2}{2} + \frac{|\alpha|^2}{2} \left
  \langle \left(X_c\left(\varphi + \frac{\pi}{2}\right) - X_d\left(\varphi
  + \frac{\pi}{2}\right)\right)^2 \right \rangle.
\end{equation}
This formula reduces to formula (\ref{delta_S_x}) giving the same
condition for non-separability of outcoming EPR mixed state.

\section{Conclusions}

We have proposed a measurement of the uncertainties of Stokes
parameter $\bar{S} = (S_x, S_y, S_z)$ as an entanglement witness for a
general class of EPR mixed correlated states. This method seems to be
relatively easy to implement experimentally.  If the quantum
fluctuations of the Stokes paramaters is below the shot noise (the
value obtained for a vacuum state), the state is entangled i.e.,
non-separable.

\section*{Acknowledgements}
This work was partially supported by a~MEN Grant No.~1 PO3B 137 30.

\vfill\eject
\end{document}